\documentclass[sigconf]{acmart}

\usepackage{graphicx}
\usepackage{textcomp}
\usepackage{bmpsize}
\usepackage{xcolor}
\usepackage{lipsum}
\usepackage{algorithm}
\usepackage{xspace}
\usepackage[noend]{algpseudocode}
\usepackage{epsfig}
\usepackage{epstopdf}
\usepackage{colortbl}
\usepackage{subcaption}
\usepackage{paralist}
\usepackage{booktabs}
\usepackage{verbatim}
\usepackage[export]{adjustbox}
\usepackage{mdframed}
\usepackage[font=small,labelfont=bf]{caption}
\usepackage{tikz}

\usepackage{algpseudocode}
\usepackage{array}
\usepackage{scalerel,stackengine}
\usepackage{enumitem}
\usepackage{cleveref}

\newcommand{\eat}[1]{{}}

\definecolor{ao(english)}{rgb}{0.0, 0.5, 0.0}
\definecolor{mypink1}{rgb}{1, 0.0, 0.6}

\newcommand*\circled[1]{\kern-2.5em%
  \put(0,4){\color{blue}\circle*{10}}\put(0,4){\circle{3}}%
  \put(-3,0){\color{white}\bfseries\large#1}~~}

\ifpdf
  \DeclareGraphicsExtensions{.pdf,.png,.jpg}
\else
  \DeclareGraphicsExtensions{.eps}
\fi

\stackMath
\newcommand\reallywidehat[1]{%
\savestack{\tmpbox}{\stretchto{%
  \scaleto{%
    \scalerel*[\widthof{\ensuremath{#1}}]{\kern-.6pt\bigwedge\kern-.6pt}%
    {\rule[-\textheight/2]{1ex}{\textheight}}%WIDTH-LIMITED BIG WEDGE
  }{\textheight}% 
}{0.5ex}}%
\stackon[1pt]{#1}{\tmpbox}%
}

\newcommand{\ie}{i.\@\,e.\@\xspace}
\newcommand{\eg}{e.\@\,g.\@\xspace}


\copyrightyear{2022}
\acmYear{2022}
\setcopyright{acmlicensed}\acmConference[WiSec '22]{15th ACM Conference on Security and Privacy in Wireless and Mobile Networks}{May 16 - May 19 2022}{San Antonio, Texas, USA}
\acmBooktitle{15th ACM Conference on Security and Privacy in Wireless and Mobile Networks (WiSec '22), May 16 - May 19 2022, San Antonio, Texas, USA}
\acmPrice{15.00}
\acmDOI{10.1145/3395351.3399361}
\acmISBN{978-1-4503-8006-5/20/07}

\begin{document}

%\title[Multi-Objective Optimization for Secure ADS-B Sensor Placement]{Multi-Objective Optimization for Secure ADS-B Sensor Placement}
\title{Towards Security-Optimized Placement of ADS-B Sensors}% using Multi-Objective Optimization} 

\author{Ala'~Darabseh}
%\authornote{Research assistant at Center for Cyber Security}
\email{ala.darabseh@nyu.edu}
\affiliation{%
  \institution{New York University Abu Dhabi}}

\author{Christina~P{\"o}pper}
%\authornote{Assistant professor of computer science at NYU Abu Dhabi}
\email{christina.poepper@nyu.edu}
\affiliation{%
  \institution{New York University Abu Dhabi}}

\begin{abstract}
Automatic Dependent Surveillance–Broadcast (ADS-B) sensors deployed on the ground are central to observing aerial movements of aircraft. Their %oftentimes 
unsystematic placement, however, results in over-densification of sensor coverage in some areas and insufficient sensor coverage in other areas. ADS-B sensor coverage has so far been recognized and analyzed as an \emph{availability} problem; it was tackled by sensor placement optimization techniques that aim for covering large enough areas. In this paper, we demonstrate that the unsystematic placement of ADS-B sensors leads to a \emph{security} problem, since the realization and possible deployment of protective mechanisms is closely linked to aspects of redundancy in ADS-B sensor coverage. In particular, we model ADS-B sensor coverage as a multi-dimensional security problem. We then use multi-objective optimization techniques to tackle this problem and derive security-optimized near-optimal placement solutions. Our results show how the location of sensors play a significant role in reducing the success rate of attackers by providing a sufficient number of sensors within a specific geographical area to verify location claims and reducing the exposure to jamming attacks.

\end{abstract}


\maketitle

\section{Introduction}

 Following mandatory requirement since 2020 in large parts of the worldwide airspace, commercial aircraft use the ADS-B (Automatic Dependent Surveillance–Broadcast) system \cite{blythe2011ads} to broadcast messages that are received by air-traffic control (ATC) and observation networks\footnote{such as Flightradar24, FlightAware, and OpenSky Network.}. ADS-B allows an aircraft to transmit its satellite-navigation derived location, which then enables it to be tracked throughout its flight time.  The broadcast messages are received by ADS-B sensors which are typically placed arbitrarily on-ground by volunteers and system designers.   The wide benefits of ADS-B ranging from surveillance coverage, cost efficiency to environmental sustainability make it widely adopted by most commercial aircraft in Europe and North America, and it is expected to replace the radar system as a part of the Next Generation Air Transportation System (NextGen) \cite{NextGen}. 

The placement of sensors is critical for the functioning of \mbox{ADS-B}. The current deployment of ADS-B sensors often places them arbitrarily on the ground (\eg, by volunteers contributing to the OpenSky Network~\cite{schafer2014bringing}), which creates some overcrowded areas and others without any coverage. For instance, it was shown in \cite{MAVPro} that only 44\% of ADS-B messages in central Europe are received by four or more sensors, while the remaining areas are either covered by fewer sensors or not covered at all. Optimal sensor placement has typically been investigated as an availability problem: How to place the receivers such that they provide the best coverage for a geographical area \cite{Maurosensors,NSGA2BasedNodePlacementOptimization,5873254}? Placing them too close to each other leads to high and possibly unnecessary redundancy, whereas placing them too far apart may result in lost observation areas because the wireless channel is faulty and messages may get lost. %This problem can be addressed using optimal sensor placement approaches \TODO{\cite{refs on sensor placements}}.

However, the open communication of ADS-B where messages are sent without encryption and integrity protection also suffers from the risk of spoofing and jamming attacks. 
%\eat{In spoofing attacks, an attacker would  modify transmitted messages and insert own ones; whereas in jamming attacks, the attacker interferes with the communication to prevent the reception of ADS-B messages. Consequences of such attacks can lead to incorrect observations of flight paths to disaster accidents when aircraft are no longer (correctly) tracked by the ATC system. %The lack of security due to these attempts may make disaster accidents as the aircraft will no longer be tracked by the ATC system.
%} 
To counter this risk, attack detection and prevention techniques have been deployed or developed for the ADS-B context: (1) A common approach to detect tampering with ADS-B messages and the reported aircraft locations is the use of Multilateration (MLAT) \cite{Mantilla-Gaviria2015}, which allows to validate the reported locations in the messages with  computed locations based on measurements from multiple sensors. MLAT requires the message to be received by four (or more) receivers on-ground in order to  derive the aircraft's location. (2) Other approaches for validity checking rely on a smaller number of received messages. % to provide a fine level of location checks with a lower cost (minimum number of receivers). 
Strohmeier et al.~\cite{Strohmeier2015LightweightLV} have proposed a lightweight approach to validate the received aircraft location claim based on the K-Nearest Neighbors algorithm (K-NN), requiring only two sensors for location verification in two dimensions (2D). This approach leads to less accuracy than MLAT, but can be used in geographic areas covered by only two sensors.  %While the accuracy of lightweight validation does not beat MLAT's accuracy.
%However, the accuracy of lightweight validation cannot beat MLAT accuracy, but on the other hand, the cost of required sensors is less than MLAT, which may make it a good option for system designers with minimum budget, or even, provide a security check in case of  the receivers becomes out of service at any point of time. 
(3) To counter jamming attacks, finally, it is beneficial to place sensors in maximum distances to decrease the probability that all sensors in the reception range of a certain geographic area are impacted by a jamming attack. 

The aim for an optimized placement of ADS-B sensors that does not only target coverage (\ie, availability of the ADS-B service), but supports the deployment of defense mechanisms (\ie, considers the security aspect of sensor deployment) creates the following challenge:  \emph{How can the different security requirements on the numbers and locations of ADS-B sensors best be fulfilled}? % under the constraint that full geographic areas should be covered by the ADS-B service}? %Some of them may even be contradictory. For instance, ...
We note that MLAT does not only need messages from at least four sensors, but its accuracy also relies on the level of Dilution of Precision (GDOP) \cite{zhu1992calculation} which again depends  on the locations and number of sensors.  %In addition, if the jammer tries to interfere in an area then all the crowded sensors will be affected and the area will be out of service.
From a practicality aspect, we further need to consider that real-world deployments cannot easily be modified to reflect optimal sensor placements. This creates a second challenge: \emph{Knowing what an optimal sensor placement from the security perspective would be, how can existing placements of ADS-B sensors best be enhanced by placing a certain number of additional sensors in a given area?}

%Regardless to which security check the system designers decide to adopt, they should know the best location of $n$ available sensors in a way that guarantees each message to be received by ADS-B receiver and one of the security checks can be applied. Moreover, the sensors are vulnerable to jamming attacks, the attacker can jam the communications and prevent the sensors from receiving the ligament messages from aircraft,  the more sensors are available in the jammer area the more victims are infected by this attack.  For this reason, the location of receivers on-ground matters to reduce the effect of the jamming attacks. 
 
%The current deployment of receivers lacks of full surveillance coverage and is not dealing with these challenges. More sophisticated sensor placement strategy is required to solve the Optimal Sensor Placement (OSP) problem.   

As a reaction to these questions, we tackle the problem of \emph{Optimal Sensor Placement (OSP) for ADS-B from the security perspective}, incorporating the constraints imposed by attack detection techniques. In particular, we define a multi-objective optimization (MOOP) problem and propose a set of solutions that satisfy these objectives simultaneously. We tackle this problem with respect to three security dimensions: First, we consider MLAT for verifying  aircraft locations claims in the received ADS-B messages. %, the requirements, and the challenges for this check. 
In this objective, we aim to provide a sensor placement solution where each broadcast message is to be received by at least four sensors and can  accordingly be verified by an MLAT check.  Second, since the cost of MLAT checks is relatively high, and it requires at least four sensors, which is hard to guarantee, we introduce a second objective: Location verification checks at a lower cost level but with less accuracy compared to MLAT. Finally, in our third objective, we aim to provide a sensor setup that behaves favorably under jamming attacks. We discuss three directions that can potentially reduce the effect of jamming attacks, (1) Maximize the distance between the sensors, (2) Maximize the distance between the jammer and the sensors, (3) Minimize the number of sensors within therange of the jammer. Thus the solution should be optimized concerning to all of these dimensions.

We treat each security dimension as an objective function to be optimized. Our target is to solve the OSP problem by providing solutions for sensor coverage that allow the aircraft to be tracked during their flight time while allowing the deployment of security checks that enable to verify ADS-B messages and place sensors in a way that mitigates the effect of a jamming attack. The main key of our approach is that all objective functions are optimized simultaneously, where each solution is non-dominated by another one. For this purpose, we adopt the Non-dominated Sorting Genetic Algorithm (NSGA-II) \cite{NSGAII} algorithm.

%\subsection{\textbf{Main Contributions}}

In short, our main contributions in this paper are:
\begin{compactitem}
\item We model the problem of ADS-B sensor placement under security considerations as a Multi-Objective Optimal Sensor Placement Problem (OSP) with respect to three concrete security objectives---two for location verification (MLAT \& leightweight verification) and one for jamming prevention.
%\item We specify and formally introduce optimal sensor placement approaches for MLAT-based location verification of ADS-B messages and reducing the effect of jamming attacks by placing ADS-B sensors in a systematic way that considers three security directions which affect the success of jamming attacks.
\item We specify and formally introduce the three security objectives and investigate their impact on ADS-B sensor locations. For the case of jamming attacks, our approach consists of a systematic way of including three security directions that impact the success of jamming attacks.  
%\item Define the Multi-Objectives Optimal Sensors Placement Problem (OSP) to localize the sensors on-ground with respect to three concrete security objectives. 
\item We provide a set of non-dominated solutions for the proposed problem. Each solution provides a sensor placement with respect to the defined objective.%, where the system designer can choose based on their budget and security levels that they are satisfied with. 
%location-verification and jamming-prevention MLAT and Lightweight location verification security checks
\end{compactitem}

\section{Preliminaries}
\label{Sec_prelim}

\subsection{\textbf{Problem Statement}}

In this paper, we address the following two research questions:
\begin{compactenum}
\item \textbf{Optimal Setting}: How can we determine the minimum number of ADS-B sensors and their near-optimal locations that are required to cover a specific geographic area while supporting the deployment of security mechanisms, in particular  location-verification checks and jamming-prevention techniques?  
\item \textbf{Real-World Setting}: For a geographic area containing already deployed ADS-B sensors (with possibly insufficient coverage), how can we find the minimum number of new sensors and their locations that should be added to the existing sensors to reach a close-to-optimal sensor deployment?
\end{compactenum}

\subsection{\textbf{Air-to-Ground ADS-B Signal Propagation}}

For the transmitted message to be received by a receiver, the  Line-of-Sight (LOS) condition must be met.  There are several factors with affect on the signal reception such as the transmitter power, the antenna gain of both transmitter and receiver, the distance between transmitter and receiver, Earth curvature, and environmental obstacles such as mountains and high buildings. The maximum message reception range corresponds to the radio horizon \cite{schafer2014bringing}  and under the  tropospheric refraction, is given by  \cite{strohbehn1968line}:
\begin{equation}
\label{rh}
\nonumber
r_0 = 3.57 \sqrt{\mathit{k}_e}(\sqrt{h_1} + \sqrt{h_2})
\end{equation}
where $\mathit{k}_e$ represents  the effective earth-radius factor and $h_1$ (resp.~$h_2$)  the transmitter (resp.~receiver) antenna height in meter. Through linear approximation of the refractivity gradient, it is shown in \cite{blaunstein2007radio} that $\mathit{k}_e \approx 4/3$.

%More details are explained in Appendix \ref{TDOA-appendix}.

If the receiver is located   beyond the  radio horizon of the transmitter, it will not receive the transmitted message and we say it is located out of the reach of the transmitter. It follows that  from Equation \ref{rh} in order for the broadcast message from a transmitter $\mathbf{p}$ to be received by the sensor $\mathbf{s}$, the following condition must be met:
\begin{equation}
\label{visi}
h_1 \ge 0.0785 \frac{\lVert \mathbf{p}-\mathbf{s} \rVert ^2}{\mathit{k}_e}.
\end{equation}

\subsection{\textbf{Time-of-Arrival Localization}}
Time-of-Arrival (TOA) is one of the most widely adopted techniques  to localize objects in radar systems, wireless sensor networks, and Internet of Things environments. The idea of this method is to determine the location of a transmitter using the time of arrival of broadcast messages by multiple, typically four, receivers. In more details, assume that there are four distributed receivers at locations  $ \mathbf{s_i} = [x_i, y_i, z_i]^T, i = 1,\ldots,4$. The TOA $t_i$ of a signal sent from a transmitter at location $ \mathbf{p} = [x, y, z]^T$ to the $i^{th}$ sensor is given by~\cite{Ichen2009wireless}: 
\begin{equation}
\label{toa}
t_i = \frac{1}{c} \lVert \mathbf{p}-\mathbf{s_i} \rVert + \tau + e_i,
\end{equation}
\noindent where $\tau$ is the signal transmission time, $c$ the speed of light, and $e_i$ the measurement error.

The location estimation of the transmitter $\mathbf{p}$ can be derived by solving the above formula of three TOA differences.  There are some factors that affect the accuracy of the obtained results, like the transmission range error and location of the sensors. MLAT is an example that has been used to localize aircraft using TDOA.

\subsection{\textbf{Geometric Dilution of Precision (GDOP)}}

To improve the accuracy of MLAT-based aircraft location estimation, the impact of the geometric dilution of precision (GDOP) should be controlled or decreased. The GDOP is affected by the geometry of locations of the receiving sensors: Placing the receivers close to each other will increase  the GDOP value and consequently degrade the accuracy of MLAT, while the accuracy will be improved if the receivers are placed far away from another\eat{ since then the GDOP is degraded}. The GDOP is defined as \cite{zhu1992calculation}: 
\begin{equation}
\label{gdop}
GDOP = \sqrt{tr(B^TB)^{-1}}
\end{equation}
where 
\begin{equation}
\label{B}
B  = \big \{[b_{i1}, b_{i2}, b_{i3}, 1] \big \}_{i = 1, \dots, 4}
\end{equation}
and $b_{i1}, b_{i2}, b_{i3}$ are the direction cosines from the aircraft to the $i^{th}$ sensor and $tr$ is the trace of the GDOP matrix. Assuming independent range measurement errors with equal variance $\sigma^2$, it holds \cite{manolakis1996efficient}:
\begin{equation}
\label{cov}
GDOP = \frac{\sqrt{tr(\mathbf{P})}}{\sigma},
\end{equation}
where $\mathbf{P}$ is the covariance matrix of the estimation error on the transponder location. To compute the GDOP value, we must derive the matrix $B$ defined in Equation (\ref{B}) that one would obtain from estimating the location of the transmitter aircraft $ \mathbf{p} = [x, y, z]^T$ using the measured ranges of four receivers $ \mathbf{s_i} = [x_i, y_i, z_i]^T$ in the ECEF (Earth-Centered, Earth-Fixed) coordinate system.  %More details are defined in Appendix \ref{GDOP-Appendix}.  
If $\phi$ and $\lambda$ are   the geodetic latitude and longitude of $\mathbf{p} = [x, y, z]^T$ respectively, then the vector $\mathbf{v_i}$ pointing from the aircraft to the $i^{th}$  sensor in the North-East-Down (NED) frame is given by:
\begin{equation}
\label{ned}
\mathbf{v_i} = [\alpha_i, \beta_i, \gamma_i]^T  = R \cdot (\mathbf{s_i} - \mathbf{p} )
\end{equation}
where the rotation matrix $R$ is :
\[
R = 
\begin{bmatrix}
    -sin(\phi) cos(\lambda)  & -sin(\phi)sin(\lambda) & cos(\phi)\\
    -sin(\lambda)& cos(\lambda) & 0 \\
   -cos(\phi)cos(\lambda)& -cos(\phi)sin(\lambda) & -sin(\phi)
\end{bmatrix}
\]

It follows that: 
\begin{equation}
\label{bcomp}
b_{i1}= \frac{\alpha_i}{\lVert \mathbf{v_i} \rVert}; \quad b_{i2}= \frac{\beta_i}{\lVert \mathbf{v_i} \rVert} ; \quad b_{i3}= \frac{\gamma_i}{\lVert \mathbf{v_i} \rVert}
\end{equation}

The GDOP at location $\mathbf{p}$ can subsequently be computed from Equation (\ref{gdop}) using the closed-form expression derived in \cite{zhu1992calculation}.

%\subsection{\textbf{ Lightweight Location Estimation Principle}}

%\subsection{\textbf{Jamming-to-Signal Ratio (JSR)}}
%\subsection{\textbf{Sensor Deployment Optimality Criteria}}

\subsection{\textbf{Genetic Algorithm}}

\label{NSGA}
%A number of methods and algorithms can be used to solve multi-objective function optimization problems. 
Traditional methods for solving multi-objective function optimization problems scalarize all the objectives into one objective using a weight vector \cite{NSGA_old}. In such a scalarization process, the obtained solution depends on the weights that are specified by the user  and it can be highly sensitive to this weight vector; the user needs to have prior knowledge of the problem in order to specify suitable weights. 
Moreover, such methods derive one possible solution only, whereas it may be interesting to get more than one solution. 
Genetic Algorithms (GA) \cite{GA_Katoch2021ARO} %comes as one of the best algorithms to 
solve Multi-Objective Optimization Problems (MOOP) by providing a set of Pareto-optimal solutions. Although the GA gives encouraging results, it shows a level of bias towards some regions since the set of solutions that are given by GA they are not disrupted fairly over the specified area where the way that select the population does not do fairly random selection of candidates \cite{GA_thesis}. 

Non-dominated Sorting Genetic Algorithm (NSGA) \cite{NSGA_old} has been proposed by Srinivas \& Deb to solve this issue and eliminate the bias by distributing the population over the entire Pareto-optimal region. However, as demonstrated in \cite{NSGA_996017}, issues with NSGA include computation complexity and lack of elitism. Thus, NSGA-II \cite{NSGA_996017} has been proposed to solve these issues and to provide solutions for MOOP. NSGA-II is well known to be a elite-preserving, fast sorting multi-objective genetic algorithm. Unlike other optimization algorithms, NSGA-II optimizes all objectives simultaneously, where each objective solution is non-dominated by other objective solutions.  

Crowding Distancing \cite{YANG2014197} of NSGA-II reflects how far the solution is from the solution boundary, so if there are two solutions, the solution with a better rank will be selected; however, if these two solutions have the same rank then the solution is selected based on its crowding distance. Thus, the key features of applying the elitist, and the attention of using the non-dominated solutions encourage us to adopt NSGA-II to solve our OPS problem.  

\begin{comment}

\begin{figure}[tbp]
\centering
{\includegraphics[width=.75\linewidth]{Illustrations/Overview.png}}
\caption{System Overview: We show how the distribution of sensors affects the message reception and the jammer success rate.}
\label{fig:system_overview}
\end{figure}
\end{comment}

\section{Threat Model}
\label{attackmodels}

We consider two types of attacks with impact on ADS-B communication:% that our proposed approach mitigates by an optimized new sensor placement strategy: 
\begin{compactenum}
    \item \textbf{ADS-B Location Spoofing}: The attacker exploits the open nature of ADS-B communication to modify the content of transmitted ADS-B message that are within the attacker's range. He/she can insert own messages or modify the broadcast location of the aircraft which leads to receiving a modified location by the on-ground sensors. The success of the attack depends on whether the attacker is located in a geographical area that is only covered by few sensors, where location verification methods cannot be applied. %\eat{Thus, our target is to locate the sensors in a way that one of these verification methods can be used to detect such type of attacks.}
    \item \textbf{ADS-B Jamming}: The attacker tries to block the communication that is received by the on-ground sensors by causing interference on the wireless channel to prevent the reception of ADS-B messages. The attacker can use any software radios with amplifiers that are typically cheap and affordable. %\Ala{In particular, we are not testing specific jammer configuration, while we only consider the dimensions that could potentially reduce the effect of jammer. As long as, these dimensions are optimized to the optimal the solution will be resilient against any powerful jammer.}
    %\eat{However, in this paper we are preventing the jamming attack while we are mitigating its affect by reducing the number of sensors the are within the range of the attacker.} 
\end{compactenum}
Consequences of such attacks can lead to incorrect observations of flight paths to disaster accidents when aircraft are no longer (correctly) tracked by the ATC system.   

\section{Overview: Objective Functions}
\label{Sec:OFunctions}
We aim to solve the Optimal Sensor Placement (OSP) problem under the consideration of multi-objective functions (MOF) that provides defense mechanisms in addition to full coverage. We use the NSGA-II algorithm \cite{NSGA_996017} to derive the best solution that satisfies all required objectives as best as possible in order to place the sensors on ground.  Since we are dealing with multi-objective functions, the NSGA-II may derive more than one solution, where each solution reflects the trade-off between the corresponding coverage and security level of all objectives.

%of these objectives and the corresponding coverage and security level of each solution.  
We tackle the OSP problem from two dimensions: First, we consider a geographical area without ADS-B receiver coverage in an idealized scenario, where we aim to derive an optimal distribution of the sensors from scratch as an upper bound of how good the best solution can be. Second, we take real-world considerations into account and consider a geographical area where sensors are deployed already but their number and locations do not provide an optimal placement -- we aim to identify how close this scenario can come to the idealized optimal solution when new sensors are added. \eat{The optimal solution here is relative and depends on the system specifications such as the budget and security level that the system designer would like to adopt.} \eat{ Figure \ref{fig:system_overview} represents an overview of the proposed system objectives and the problems we are targeting: Some broadcast messages are not received by any sensors whereas other messages are received by a lot of sensors. Also, it shows how the location of the sensors affects the number of jammed sensors by the attacker. }

Our approach provides several solutions with trade-offs as will be  later shown in Section~\ref{sec:experiments}.  Here, we next describe the set of  security objectives that these derived solutions are based on; for each objective we provide a detailed specifications and security level.

\subsection{\textbf{Objective Function 1: Multilateration (MLAT) surveillance under GDOP}}
\label{OF1_GDOP}

MLAT is a standard technology to localize aircraft or verify the received location from aircraft. Whenever we say the airspace supports MLAT checks we refer to verifiable location claims. %Also, the level of security this method can provide is very high and accurate. 
Since MLAT requires four or more sensors to receive a message, %Such a requirement is hard to achieve with the deployment of the current sensors, where there are some areas (around 70\%) are only covered by three sensors or less. 
by this first objective, we will identify the best sensors deployment solution such that each message can be received by at least 4 sensors on the ground. 

Let us assume an airspace $\mathcal{A}$ contains the expected ADS-B traffic. Then we take $m$ sample locations $\mathbf{p_j}$ from the  $\mathcal{A}$ volume space:
\begin{equation}
%\label{spat_A}
\nonumber
\label{Equ-1}
 \mathcal{A} =  \{g_j | \mathbf{p_j}\}_{j=1,\cdots,m}
\end{equation}
where, $g_j$ represents the \textit{required} GDOP value at $\mathbf{p_j}$.  In addition, given a placement $\mathbf{S} = \{ \mathbf{s_i}\}_{i=1,\cdots,n}  $ of $n$ ADS-B sensors, we write $\reallywidehat{g_j}=gdop_{\scaleto{\mathbf{S}}{4pt}}(\mathbf{p_j})$ to denote the \textit{achieved} GDOP value at the location $\mathbf{p_j}$ due to the particular geometry of $\mathbf{S}$. For readability, we omit the subscript and write $\reallywidehat{g_j}=gdop(\mathbf{p_j})$.

In order to find the best deployment of ADS-B sensors and their corresponding locations from $\mathbf{S}$, to satisfy a per-location GDOP requirement for a given airspace $\mathcal{A}$, we assume $\delta$ to be the tolerance parameter on the GDOP at any location, where $\mathbf{S}$ can only be accepted if: 
\begin{equation}
%\label{cond}
\nonumber
\label{Equ-2}
\forall \;  \mathbf{p_j} \in \mathcal{A}, \; \;  |\reallywidehat{g_j} - g_j| < \delta
\end{equation}
which is equivalent to $\lVert \mathbf{\reallywidehat {g}} -  \mathbf{g} \rVert_{\infty} < \delta$, where $\lVert . \rVert_{\infty}$ represents  the sup-norm defined by
%\begin{equation}
%\label{sup2}
%\nonumber
%\label{Equ-4}
$\lVert \mathbf{\reallywidehat {g}} -  \mathbf{g} \rVert_{\infty} = \max_{j}{|\reallywidehat{g_j} - g_j|.}$
%\end{equation}

The Mean Squared Deviation (MSD)  between the achieved and the required GDOP in the entire airspace of our first objective is:
\begin{equation}
%\label{msd}
\label{Equ-5}
  MSD(\mathbf{S}) = \frac{1}{m}\sum_{j=1}^{m} (g_j - \reallywidehat{g_j})^2.
\end{equation}

\subsection{\textbf{Objective Function 2: Lightweight Location Check with Transmission Range Evaluation}}
\label{OF2_Transmission_range}
As described in Section \ref{Sec_prelim}, the Lightweight Location Estimation~\cite{Strohmeier2015LightweightLV} is based on TDOA, where the received message is received by at least two sensors. %While TDOA is restricted in indoor environments due to signal multi-path reflections, still it can provide promising performance for outdoor environments \cite{strohmeier2018k}.  
We can thus design our second objective to provide another security check with fewer sensors: only two sensors are required. However, the accuracy of this check must be assumed to be less than for the MLAT-check (objective~1), but still, this lightweight method can provide fine results with small budget, and it can be used in areas where MLAT is not available either due to lack of a number of sensors or an attack that affects the area and disrupts some of the sensors.   

To achieve the second objective, we use the transmission range or distance as an evaluation principle to deploy the sensors.  In more details, given an airspace $\mathcal{A}$ with    $m$ uniformly sample locations $\mathbf{p_j} $ from $\mathcal{A}$,  the following spatial data matrix is defined:
\begin{equation}
%\label{spat_A}
\label{Equ-6}
\nonumber
 \mathbf{A} =  \{tr_{(j,s)} | \mathbf{p_j}\}_{j=1,\cdots,m}
\end{equation}
where, $tr_{(j,s)}$ represents the \textit{best} minimal distance  from $\mathbf{p_j}$ to sensor  $\mathbf{s_i}$, the required one.  In addition, given a placement $\mathbf{S} = \{ \mathbf{s_i}\}_{i=1,\cdots,n}  $ of $n$ ADS-B sensors, we write $\reallywidehat{tr_j}=tr_{\scaleto{\mathbf{S}}{4pt}}(\mathbf{p_j})$ to denote the \textit{achieved} distance from the location $\mathbf{p_j}$ to $ \mathbf{s_i}$ due to the particular geometry of $\mathbf{S}$. For readability, we omit the subscript and note $\reallywidehat{tr_j}=tr(\mathbf{p_j})$. 

Now, to find the minimal number of ADS-B sensors $n_{min}$ and their corresponding deployment $\mathbf{S}$ to guarantee a per-defined transmission range requirement for a given airspace $\mathcal{A}$,  let us assume $tr_\delta$ be the tolerance parameter of $tr$ at any location. A sensor placement  $\mathbf{S}$ can only be accepted if:
\begin{equation}
\label{Equ-7}
\nonumber
\forall \;  \mathbf{p_j} \in \mathcal{A}, \; \;  \lVert \mathbf{\reallywidehat {tr}} -  \mathbf{tr} \rVert_{\infty} < tr_\delta
\end{equation}

where $\lVert . \rVert_{\infty}$ represents  the sup-norm defined by:
\begin{equation}
\label{Equ-9}
\nonumber
\lVert \mathbf{\reallywidehat {tr}} -  \mathbf{tr} \rVert_{\infty} = \max_{(j,s)}{|\reallywidehat{tr_{(j,s)}} - tr_{(j,s)}| }
\end{equation}

The Mean Squared Deviation (MSD) between the achieved and the required $tr$ in the entire airspace under consideration is
\begin{equation}
\label{Equ-10}
  MSD(\mathbf{S}) = \frac{1}{m}\sum_{j=1}^{m} (tr_{(j,s)} - \reallywidehat{tr_{(j,s)}})^2.
\end{equation}

\subsection{\textbf{OF~3: Low Sensor Density under Jamming}}
\label{OF3_Sensor_Density}

Tackling jamming attacks requires more sophisticated considerations to optimize the locations of  sensors on ground. The aim of this objective is to reduce the effect of jamming by placing the sensors in a way that guarantees good coverage, while keeping the number of sensors affected by the jammer to a minimum.  We incorporate three directions for the network topology of deployed sensors:

\begin{compactitem}
    \item \textbf{Direction 1:} Maximize the distance between any two sensors $dist(s_i,s_{\overline{i}})$, where $s_i$ and $s_{\overline{i}}$ are any two sensors in $\mathbf{S}$ .
   The aim is to select the best candidates of sensors that are placed in %a way where each two sensors
   locations far away from each other, in other words, the best low sensor density network. 
   
   In more details, given the an airspace  $\mathcal{A}$ like in the first two objectives, we need to fully cover it by a set $\mathbf{S} = \{ \mathbf{s_i}\}_{i=1,\cdots,n}  $ of $n$ ADS-B sensors. Now, to find the minimal number of ADS-B sensors $n_{min}$ and their corresponding deployment $\mathbf{S}$ to guarantee full coverage for given airspace $\mathcal{A}$ and low sensor density,  let us assume $dist_\delta$ be the tolerance parameter of $dist$ between any two sensors. A sensor placement  $\mathbf{S}$ can only be accepted if:
\begin{equation}
\label{Equ-11}
\nonumber
\forall \;  \mathbf{s_i},\mathbf{s_{\overline{i}}}\in \mathcal{S}, \; \;  \lVert \mathbf{ \reallywidehat{dist_{(s_i,s_{\overline{i}})}}} -  \mathbf{dist} \rVert_{\infty} > dist_\delta
\end{equation}

where $\lVert . \rVert_{\infty}$ represents  the sup-norm defined by:
\begin{equation}
\label{Equ-13}
\nonumber
\lVert \mathbf{ \reallywidehat{dist_{(s_i,s_{\overline{i}})}}} -  \mathbf{dist} \rVert_{\infty} = \max_{(s_i,s_{\overline{i}})}{|\reallywidehat{dist_{(s_i,s_{\overline{i}})}} - dist| }
\end{equation}

The Mean Squared Deviation (MSD) of the distance between any two sensors under consideration is:
\begin{equation}
\label{Equ-14}
  MSD(\mathbf{S}) = \frac{1}{n} \sum_{i=1}^{n} (dist_{(s_i,s_{\overline{i}})} - \reallywidehat{dist_{(s_i,s_{\overline{i}})}})^2
\end{equation}

 %%%%%%%%%%%%%%%%%%%%%% Direction 2  
    \item \textbf{Direction 2:} Maximize the distance between the jammer and the sensors $dist(jam_i,s_j)$.
    This direction aims to reduce the Jamming-to-Signal (JSR) ratio \cite{manuals}. 
    \begin{equation}
    \label{Equ-15}
        JSR= \frac{P_jG_j{dist_{(t,s)}}^2}{P_TG_T{dist_{(jam,s)}}^2}
    \end{equation}
    where, $P_j$ and $G_j$ are the transmission power and antenna gain of the jammer, the $P_T$ and $G_T$ the transmission power and antenna gain of the transmitter, and the $dist_{(t,s)}$, $dist_{(jam,s)}$ are the distance from the transmitter to the sensor, and the distance from the jammer to the sensor respectively.

    As we can see from the Eq.~\ref{Equ-15}, we can reduce the ratio either by changing the transmitter characteristics which is here the ADS-B out-device on all aircraft, or maximize the distance between the jammer and the receiver. Since it is hard to change the already deployed transmitters, we can work on the distance between the jammer and the receiver. 
    
    Given list of $k$ jamming attacks at different locations $\mathbf{J} = \{ \mathbf{j_l}\}_{l=1,\cdots,k}  $ within the airspace  $\mathcal{A}$,  let us assume $dist_{\overline{\delta}}$ be the tolerance parameter of $dist$ between and jammer and any sensor. A sensor placement  $\mathbf{S}$ can only be accepted if:
    
    \begin{equation}
\label{Equ-16}
\nonumber
\forall \;  \mathbf{jam_l} \in \mathcal{J} , \;  \; \forall \mathbf{s_i} \in \mathcal{S} \; \;  \lVert \mathbf{ \reallywidehat{dist_{(jam_l,s_i)}}} -  \mathbf{dist} \rVert_{\infty} > dist_{\overline{\delta}}
\end{equation}

where $\lVert . \rVert_{\infty}$ represents  the sup-norm defined by:
\begin{equation}
\nonumber
\label{Equ-18}
\lVert \mathbf{ \reallywidehat{dist_{(jam_l,s_i)}}} -  \mathbf{dist} \rVert_{\infty} = \max_{(i,l)}{|\reallywidehat{dist_{(jam_l,s_i)}} - dist| }
\end{equation}

The Mean Squared Deviation (MSD) of the distance between any jammer and any sensor under consideration is written:
\begin{equation}
\label{Equ-19}
  MSD(\mathbf{S}) = \frac{1}{k}\sum_{l=1}^{k}\sum_{i=1}^{n} (dist_{(jam_l,s_i)} - \reallywidehat{dist_{(jam_l,s_i)}})^2
\end{equation}

%%%%%%%%%%%%%%%%%%%%%% Direction 3
    \item \textbf{Direction 3:} Minimize the number of sensors within the range of the jammer.   As jammer tries to interfere with all the sensors that are within its range. We aim by this objective to reduce the number of sensors that are within the range of the jammer, where at anytime the number of affected sensors is minimal as possible.   
    
    In more details, given the airspace $\mathcal{A}$, list of sensors $\mathcal{S}$, and list of jamming attacks $\mathcal{J}$, with the assumption that  $LOS_\delta$ is the tolerance parameter of the required number of sensors within the jammer range. A sensor placement  $\mathbf{S}$ can only be accepted if:
 \begin{equation}
\label{Equ-20}
\nonumber
\forall \;  \mathbf{jam_l} \in \mathcal{J} ,  \; \;  |\reallywidehat{LOS_{jam_l}} - LOS_{jam_l}| < LOS_\delta
\end{equation}

which is equivalent to:

\begin{equation}
\label{Equ-21}
\nonumber
\lVert \mathbf{ \reallywidehat{LOS_{jam_l}}} -  \mathbf{LOS} \rVert_{\infty} = \max_{(i,l)}{|\reallywidehat{LOS_{jam_{i,l)}}} - LOS_{jam_{(i,l)}}| }
\end{equation}

The Mean Squared Deviation (MSD) of the achieved and required number of sensors within the LoS of the jammer under consideration is written:
\begin{equation}
\label{Equ-22}
  MSD(\mathbf{S}) = \frac{1}{k}\sum_{l=1}^{k} (LOS_{jam_l} - \reallywidehat{LOS_{jam_l}})^2
\end{equation}

\end{compactitem}

\begin{comment}

\subsection{\textbf{Objective Function 4: MAVPro surveillance system with Full Coverage Evaluation Principle to avoid DoS attack.}}
\label{OF4_MAVpro_cable}
\textbf{TODO....TODO....TODO....TODO....TODO....}

The aim of this objective is to place the sensors in a way that guarantees full coverage, where each ASD-B message receives by at least one sensor, to avoid DoS attack. Hence, since the two security checks in our first objectives can not be used in such deployment, there is another security check was proposed to verify the trustworthiness of received ADS-B messages, called MAVPro \cite{MAVProDara}, such security check requires only one sensor to receive the message, however, compared to MLAT, MAVPro can provide very good accuracy within a short distance, while this accuracy is reduced and affected by long-distance where MLAT does not. Also, it can verify the location in 3D dimensions, while the lightweight location in our second objective verifies the messages in 2D dimensions.   

In this objective, we design a secure lightweight sensors deployment solution, that can provide a fine level of security with a restricted budget; only a few sensors are available to be deployed so we should know where to place them.
\end{comment}

\section{\textbf{System Approach and Methodology}}
\label{approach}

%To solve the OSP problems we adopt the MSD function to get the minimum GDOP for each air point based on the sensors placement setup. More specifically, for each transmitter location $\mathbf{p_j} = [x_j, y_j, z_j]^T$ in the  ECEF frame, the corresponding geodetic coordinates are denoted $[\lambda_j, \phi_j, h_j ]^T$.  Let $\mathbf{p'_j} $ be the point on the geoid surface that has the same longitude and latitude  as $\mathbf{p_j} $. Assuming a ground-based deployment of receivers (i.e., each has an altitude of less than 10 m), it follows from Equation (\ref{rh}) that a sensor $ \mathbf{s_i}$ is in direct visibility with an emitter located at $\mathbf{p_j} $ if and only if the following inequality is satisfied:
\subsection{\textbf{Assumptions}}
For finding the optimal placement locations of sensors we make the following simplifying assumptions:
\begin{itemize}
    \item Receivers are assumed to be deployed on the geoid surface. 
    \item We neglect the obstructing  effect of buildings or mountains on the ADS-B signal reception probability. 
    \item The earth curvature is as the major impediment to the direct visibility between aircraft and ground-based sensors. 
\end{itemize}
\subsection{\textbf{System Nodes Representation}}

Under the consideration that all system nodes are deployed on the geographical surface, their locations are specified by their geodetic latitude, longitude, and altitude. Accordingly, for our surface area $ \mathcal{A}$, we represent all the nodes within the latitude ($\mu$) and longitude ($\theta$) boundaries of this area.  

\begin{enumerate}[label=(\roman*)]

\item Pick $m$ uniformly distributed points $\mathcal{P}$ on the surface area $\mathcal{A}$. 
$\mathcal{P} = \{ \mathbf{p_j }\}_{j=1,\cdots,m} $, where, each point in $\mathcal{P}$ represents the potential location of aircraft in space. $\mathcal{P} =[(\theta_j, \mu_j), \cdots,  (\theta_m, \mu_m)]$.  
\item Place $n$ uniformly sensors $S$ on the specified area space $\mathcal{A}$. $ \mathcal{S} = \{ \mathbf{s_i }\}_{i=1,\cdots,n} $, where, $\mathcal{S} =[(\theta_i, \mu_i), \cdots,  (\theta_n, \mu_n)]$.

\item Place $k$ uniformly distributed jammers across the whole area $\mathcal{A}$.  $ \mathcal{J} = \{ \mathbf{j_l }\}_{l=1,\cdots,k}$, where, $\mathcal{J} =[(\theta_l, \mu_l), \cdots,  (\theta_k, \mu_k)]$.

\end{enumerate}
 
where $\theta_{j,i,l}$ and  $\mu_{j,i,l}$ are  the longitude and latitude of the ${j,i,l}^{th}$ points, sensors, jammers nodes respectively. For each set of nodes to be added uniquely, the following constraint is added: 
\begin{equation}
\label{constr1}
\nonumber
 \forall_{f=1}^{F-1}, \quad  
 \begin{cases}
   \;\;\; \theta_f \le \theta_{f+1}   \quad  \text{and}\\
    \text{if} \; \theta_f = \theta_{f+1}  \quad   \text{then}   \quad   \mu_f \le \mu_{f+1} 
  \end{cases}  
\end{equation}

where $f,F$ represent. the $j,i,l$ and $m,n,k$ respectively. 

The geoid boundaries of the set is specified with this constrain:
\begin{equation}
\label{constr2}
\nonumber
 \forall_{f=1}^{F}, \quad  
 \begin{cases}
   \theta_{low}  \le \theta_{f} \le \theta_{up}  \\
    \mu_{low}  \le \mu_{f} \le \mu_{up}  \\
  \end{cases}  
\end{equation}

where $\theta_{low}$ (resp. $\mu_{low} $) and  $\theta_{up}$ (resp. $\mu_{up}$) are  the lower and upper bounds on each node longitude (resp. latitude).

\subsection{\textbf{Fitness/Cost Function}}

\label{Sec:FitnessFunction}

The fitness function is designed according to our multi-objective optimization problem based on the MSD that we described in Section \ref{Sec:OFunctions}. The MSD computes the fitness of selected subsets of sensors from $\mathbf{S}$ that are chosen by the NSGA-II algorithm. It evaluates the placement of each sub-set of sensors by measuring the average of errors between the achieved value and the required one to get the final score. NSGA-II searches for the optimal Pareto frontier \cite{Miettinen1998NonlinearMO}, and more precisely we consider all solutions with first Pareto front, non dominated solutions. 

As anti-jamming space objective function  \ref{OF3_Sensor_Density} considers three optimization directions; two maximization problems and one minimization problem, we deal with them as one objective by using Weighted Sum Method \cite{Weighted_sum_method}. Each direction can be assigned a weight, which reflects the importance of this direction against the other two directions, and then combine them together as one score. 
\begin{equation}
\label{WSM}
  OF_3 =   \sum_{i=1}^{3}  {Fitness_{score(i)}*w_i} 
\end{equation}
where $w_i$ is the weight for objective direction $i$, and $\sum_{i=1}^{3}  w_i =1  $.

In addition, we define a cost or penalty function to increase our security. The cost function searches for the Pareto frontier of all defined objectives in Section \ref{Sec:OFunctions} versus the cardinality of the selected sensor set. Thus, we formulate our problem as a Knapsack-equivalent \cite{knapsack_problems}, where the 2D surface area $\mathcal{A}$ is split into $R$ rectangles, where each can be assigned a senor and this will be assigned a weight 1 ($w_i=1$) or ($w_i=0$) if there is no sensor is assigned to the rectangle. The target is to have a few rectangles that are assigned a sensor (minimize the number of sensors that are needed to be deployed)  while optimizing all of our objectives. We formulate the following penalty function on the selected set of sensors.  
\begin{equation}
\label{penal}
  c(\mathbf{S})= \frac{1}{2} \cdot \bigg [\frac{1}{R} \sum_{i=1}^{L}  {w_i} \bigg]^2
\end{equation}

Eventually, the weighted fitness function can subsequently be obtained from MSD of the objective functions in \ref{Sec:OFunctions} and (\ref{penal}) as:
\begin{equation}
\label{wei_obj}
 \tilde{f} (\mathbf{S})= (1-a) \cdot f(\mathbf{S}) + a \cdot c(\mathbf{S})
\end{equation}
where $a$ is the Pareto weight of the cost function $c$. 

Finally, since we are working with multi-objective functions and each objective implies different checks and consequently different unit scales, we normalize all the obtained scores by the following equation for each objective:  
\begin{equation}
 Norm\_OF= (OF_{score} - OF_{min})/(OF_{max}-OF_{min})   
\end{equation}

where, $OF_{score}$ is the obtained score from the objective function using NSGA-II, and the $OF_{max}$, $OF_{min}$ are the maximum and minimal required scores of this objective, and the best value is assigned at the end of all generations.

\subsection{\textbf{Procedure to solve  OSP Problems}}
To solve the OSP problems that we define in Section \ref{Sec:OSP_problems}, we adopt the MSD to get the scores of all defined objectives as we explained in Section \ref{Sec:OFunctions}. Before that, we compute and derive the following structures to be used through objective computations. 
\begin{enumerate}[label=(\roman*)]

\item  Compute direction cosines matrix for all points in $\mathcal{P}$ to all sensors in $\mathcal{S}$. $DC_{Pj}  = \big \{[dc_{i1}, dc_{i2}, dc_{i3}, \dots n] \big \}_{i = 1, \dots, n}$, where, $dc_{i1}, dc_{i2}, dc_{i3}$ are the direction cosines from the airspace point $j$ to the $i^{th}$  sensor.

\item Compute direction cosines matrix for all jammers in $\mathcal{J}$ to all sensors in $\mathcal{S}$. $DC_{Jl}  = \big \{[dc_{i1}, dc_{i2}, dc_{i3}, \dots, k] \big \}_{i = 1, \dots, k}$, where $dc_{i1}, dc_{i2}, dc_{i3}$ are the direction cosines from the jammer $l$ to the $i^{th}$  sensor.

\item Compute the distance from each point in $\mathcal{P}$ to all  sensors in $\mathcal{S}$ using the defined expression of Euclidean Distance. $Dist_{Pj}  = \big \{[dist_{i1}, dist_{i2}, dist_{i3}, \dots n] \big \}_{i = 1, \dots, n}$, where, $dist_{i1}, dist_{i2}, dist_{i3}$ are the distance from the airspace point $j$ to the $i^{th}$  sensor.

\item Compute the distance from each jammer in $\mathcal{J}$ to all sensors in $\mathcal{S}$. $Dist_{Jl}  = \big \{[dist_{i1}, dist_{i2}, dist_{i3}, \dots n] \big \}_{i = 1, \dots, n}$, where, $dist_{i1}, dist_{i2}, dist_{i3}$ are the distance from the jammer $l$ to the $i^{th}$  sensor.

\item Compute the distance between all sensors in $\mathcal{S}$. $Dist_{Si}  = \big \{[dist_{i1}, dist_{i2}, dist_{i3}, \dots n] \big \}_{i = 1, \dots, n}$, where $dist_{i1}, dist_{i2}, dist_{i3}$ are the distance from the sensor $i$ to the $i^{th}$  sensor.

\begin{comment}

\begin{equation}
\label{Dist1}
dist \left( j,i\right)   = \sqrt {\sum _{n=1}^{3}  \left( j_{n}-i_{n}\right)^2 } 
\end{equation}
\begin{equation}
\label{Dist2}
dist \left( l,i\right)   = \sqrt {\sum _{n=1}^{3}  \left( l_{n}-i_{n}\right)^2 } 
\end{equation}
\end{comment}

\end{enumerate}

After Preparing all the matrices, we  compute the objective functions. Each procedure is applied at every generation of NSGA-II.  
\subsubsection{Objective 1: Minimal GDOP}
\begin{enumerate}[label=(\roman*)]

\item Find the set $\mathcal{S}_{\scaleto{LOS}{4pt}}^j$ of all ADS-B receivers for which Inequality (\ref{visi}) is valid.

\item If $|\mathcal{S}_{\scaleto{LOS}{4pt}}^j| < 4$, set $\reallywidehat{g_j} = \infty$. (The GDOP cannot be evaluated if there are less than 4 sensors in LOS condition with the aircraft)
\item Otherwise, compute the GDOP at $\mathbf{p_j}$ for all 4-sized subsets of $\mathcal{S}_{\scaleto{LOS}{4pt}}^j$ using the closed-form expression proved in \cite{zhu1992calculation}. Then set $\reallywidehat{g_j}$ to the minimal value found.

\end{enumerate}

\subsubsection{Objective 2: Minimal transmission range}

\begin{enumerate}[label=(\roman*)]

\item Find the set $\mathcal{S}_{\scaleto{LOS}{4pt}}^j$ of all ADS-B receivers for which Inequality (\ref{visi}) is valid. 

\item If $|\mathcal{S}_{\scaleto{LOS}{4pt}}^j| < 2$, set $\reallywidehat{tr_j} = \infty$. (The lightweight location check can not be evaluated if the umber of sensors are less than two)

\item Get the distance from each point in $P$ to the all selected sensors from $\mathcal{S}$ from the matrix $Dist_{Pj}$.

\item Get the to the minimal two values, which represent the closest two sensors, and assign it to the $\reallywidehat{tr_j}$. 

\end{enumerate}

\subsubsection{Objective 3: Anti-Jamming area} 
%As Anti-Jamming objective has three direction, we show how is the procedure for each one separately: \\

\begin{enumerate}[label=(\roman*)]
   \item  For each sub-set of selected sensors from $\mathcal{S}$ get the distance between them from the $Dist_{Si}$
   \item Get the best sub-set that has the maximum distance from all sensors (\textbf{Direction 1}). 
   \item  For each jammer $l$ in $\mathcal{J}$, find the set $\mathcal{S}_{\scaleto{LOS}{4pt}}^l$ of all ADS-B receivers that are within the range $l^{th}$ jammer. 
   \item  If $|\mathcal{S}_{\scaleto{LOS}{4pt}}^l| = 0$, set $\reallywidehat{dist_{jam2s}}= \infty$ (\textbf{Direction 2}), and $\reallywidehat{LOS_{jam}} = 0$ (\textbf{Direction 3})  (Best solution).
   \item Otherwise, get the distance from each $l$ for all sensors from the $Dist_{Jl}$, and then get the minimum one to get the score of $\reallywidehat{dist_{jam2s}}$ , and repeat step 3 and step 4 for all sub-sets of sensors to find the $min (|\mathcal{S}_{\scaleto{LOS}{4pt}}^l| ) $ and assign it to $\reallywidehat{LOS_{jam}}$. 
   
\end{enumerate}

\section{\textbf{Selecting the Optimum}}

\label{Appendix:Opt}
 
%The optimization technique of NSGA-II, as we explained in Section \ref{NSGA}, optimizes all the objective functions simultaneously where each solution can not be dominated by any other. 

The optimization technique of NSGA-II, as we explained in Section \ref{NSGA}, optimizes all the objective functions simultaneously where each solution can not be dominated by any other. However, there is no solution that can satisfy all the objectives together.  In more detail, if there are three solutions $A$, $B$, and $C$, where $A$ and $B$ belong to the first Pareto frontier and $C$ belongs to the second one. We can say solution $A$ can not be dominated by solution $B$ since $A$ gives better solution for an example objective $f_1$ one, while $B$ gives better solution for objective $f_2$ where both of them are dominated by solution $C$ because they offer better solution for $f_1$ and $f_2$ than $C$.

Selecting the best solution depends on few factors like the Budget, the security level, level of redundancy and level of noise (Appendix \ref{Appendix:Opt} for more details). There is no one solution that satisfies all objectives. The optimal solution is relative here. More details will be explained in Section \ref{sec:experiments}.

Selecting the best solution, the best number of sensors, and their locations as in our problem depends on some factors, which include: 
\begin{enumerate}
    \item \textbf{ Budget:} Sensor placement is restricted to the number of available sensors that would have to place. More available sensors give flexibility and simplify the process of choosing the best numbers.  
    \item \textbf{The required security level:} As we explained in Section \ref{Sec:OFunctions}, there are different security checks and each one has some requirements and at the same time provides a level of accuracy, as an example MLAT is more accurate than lightweight location verification. Thus, the best solution depends on the required security level, if the system is critical then the solution which gives better scores for objective number one should be chosen.  On other hand, if the system is restricted to $n$ number of sensors and at the same time there is should be a level of security, then another solution that satisfies this requirement would be more valuable for this case. 
    \item \textbf{The level of redundancy:} Network leakage could be caused by some factors, like the sensor goes turned off because of an empty battery or any reason or by infecting it by an attack. For whatever reason, if the area is covered by only this sensor, then all broadcast messages will not be received. Taking into consideration this factor, a solution that gives a fine level of redundancy has to be chosen.  
    \item \textbf{Level of noise from crowded sensors:} Placing  many sensors close to each other may potentially produce  level of noise, and as a result reflects on MLAT accuracy. %, so another factor plays an important role in the best solution selection.  
\end{enumerate}

\begin{comment}
\subsection{\textbf{Sensors Ranking}}

As NSGA-II gives a number of best solutions for the number of sensors, we decided to check and give the sensors rank, where each value of the rank depends on the frequency of appearance of this sensor in all solutions with the first Pareto frontier. 

After ranking them, we choose the highest $n$ sensors and we get the fitness function of these set of sensors and see if such placement can give a better solution or not, or at least see how far this solution is from the Pareto front line. 
\end{comment}
\section{OSP problems and Case Studies}
\label{Sec:OSP_problems}

\subsection{\textbf{Scenario~1: OSP from Scratch}}
In this scenario, we consider the situation where the volume space $\mathcal{A}$ is uncovered by any sensor. Thus, we have to find the best minimal number of $n$ ADS-B receivers and their locations to cover it. That is, each $p_j$ in $\mathcal{A}$ is assigned the required GDOP value, $\reallywidehat{g_j}$ and the required (closest) distance from this $p_j$ to the ground, $\reallywidehat{tr_j}$. In addition, the best-required distance between the jammers and sensors, and the required distance between sensors are defined.  

In an ideal scenario, we wish to cover $\mathcal{A}$ with a budget of $n$ sensors where all objectives in Section \ref{Sec:OFunctions} are satisfied and the differences between achieved and required values of all objectives are equal to zero. Although this case is hard to achieve in practice, we optimistically look to find a sensor placement solution that is as close as possible to this ideal case.  

\subsection{\textbf{Scenario~2: Optimal Network Augmentation}}

In reality, there are already deployed ADS-B receivers on the ground. However, the current deployment does not guarantee full coverage, and as a consequence security checks can not be applied. By this scenario, we wish to augment the current deployment by adding $n^*$ additional receivers to the existing ones to provide a near-optimal solution.  We assume the current sensors network consists of $n$ deployed receivers $\mathbf{S^{depl}} = \{ \mathbf{s_i}\}_{i=1,\cdots,n}$ at known locations that are obtained from OpenSky database. We look to find the best number $n^*$ of new sensors $\mathbf{S^{new}}$ that can be added to the deployed ones $\mathbf{S^{depl}}$ to provide the best near-optimal solution.

Suppose $f$ new sensors $\mathbf{S^{new}} = \{ \mathbf{s_i}\}_{i=n+1,\cdots,n+f}$ must be deployed. Then  $\mathbf{S} = \big \{\mathbf{S^{depl}},  \mathbf{S^{new}} \big \} $. The set of  new receivers $\mathbf{S^{new}}$ are chosen from a predefined set of candidate ones $\mathbf{S^{cand}} = \{ \mathbf{s'_f}\}_{f=1,\cdots,N}$, where the locations of these candidates are already known.

%\subsection{\textbf{Scenario \#3: Best Sensors to Guarantee a  Surveillance Requirement}}

\section{Experimental Evaluation}
\label{sec:experiments}
\subsection{\textbf{First Findings and Observations ("Random Sensor Placement")}}
\begin{table}
  \caption{Fitness Values of Objectives with Random Placement of 21 sensors from OpenSky}
  \label{tab:fitness}
  \begin{tabular}{ccccl}
    \toprule
      & OF1 & OF2 & OF3   \\
    \midrule
    Fitness Value & 0.02203632 & 0.02732189 &  0.05420901   \\
    
  \bottomrule
 
\end{tabular}
 \vspace*{-5mm}
\end{table}

\begin{figure}[!t]
\centering
\includegraphics[ width=3.7in]{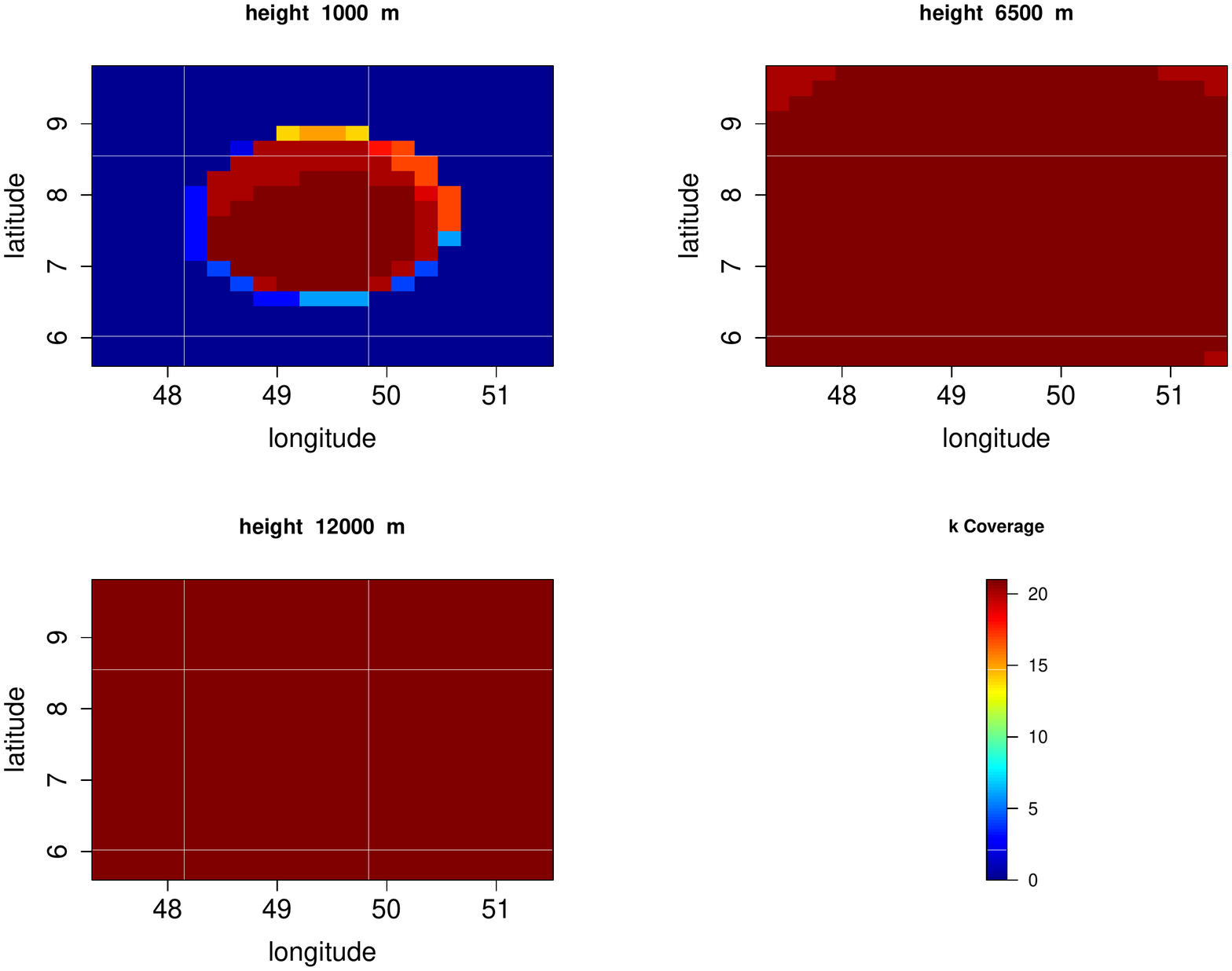}
\caption{Observed k-coverage heatmap for a random placement of $n=21$ ADS-B already deployed sensors.}
\label{fig:Kcovdep}
\vspace*{-5mm}
\end{figure}

\begin{figure}[!t]
\centering
\includegraphics[ width=3.7in]{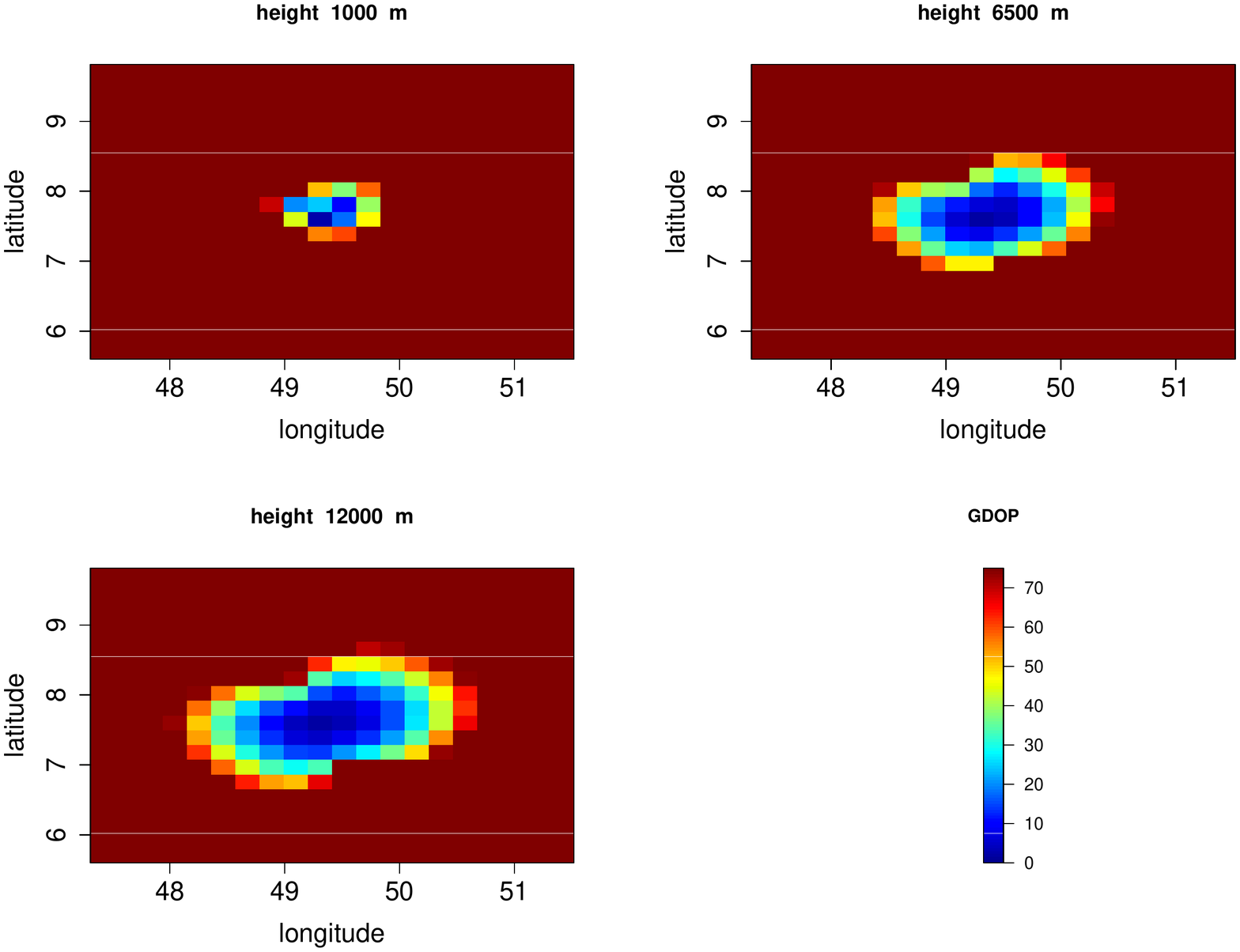}
\caption{Observed GDOP values for a random placement of $n=21$ ADS-B already deployed sensors.}
\label{fig:GDOPdep}
\vspace*{-5mm}
\end{figure}
To evaluate the effectiveness of our system and check the objective function of the proposed solutions by NSGA-II, we consider a small geographical area between 47.4 to 51.4 latitudes decimal degrees and 5.71 and 9.71 longitudes decimal degrees. We were able to get the location of 21 ADS-B receivers from OpenSky for this specified area. The current deployment of these sensors is placed randomly by the users. Thus, we evaluate the current deployment and we check the fitness values of all objective functions. We test the individual objective function and all the combinations of objectives to check how far is the current deployment to the optimal solution. Table \ref{tab:fitness} shows the fitness function values of the objective functions. We consider these values as a reference point to show how far is the current deployment from the optimal scenario. 

As our first objective targets to have full coverage while minimizing the GDOP value, we test how are the k-coverage and GDOP values of the 21 ADS-B receivers from OpenSky. Figure \ref{fig:Kcovdep} represents the heatmap of the sensor's coverage (k-coverage) of the specified area across different heights (altitude). As we can see from the figure, aircraft at low height have low coverage which limits the MLAT check during take-off and landing phases (the most critical and important phases of aircraft path), but once the aircraft goes up it will be visible (within LOS) to more sensors. However, such deployment also produces a significant amount of GDOP values which reduce the accuracy of the MLAT verification test as figure \ref{fig:GDOPdep} reflects the condense of GDOP values for this random placement.

\begin{figure}[tbp]
\centering
{\includegraphics[width=.80\linewidth]{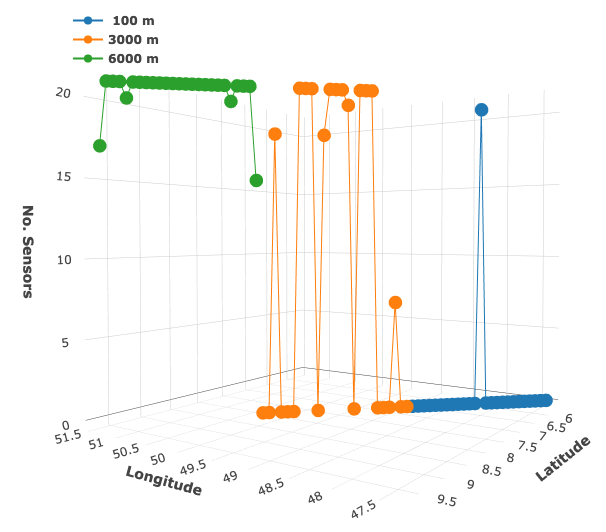}}
\vspace*{-3mm}
\caption{The number of sensors that are affected by the jamming attack setup (75 jamming attacks at three height levels across the whole area) with the $n=21$ already deployed sensors from OpenSky. Each dot in the figure represents the location of the jammer in 2D.}
\label{fig:Jamming_dep}
\vspace*{-3mm}
\end{figure}

Moreover, we evaluate the resistance of the current deployment against the jamming attack (our third objective). To test that, we generated 75 jamming attacks across the whole area at different altitudes. We assume the attacks could be at different locations, like on the high building, or even on the aircraft. We check how many sensors out of 21 sensors are affected by such a setup. Figure \ref{fig:Jamming_dep} reflects how successfully the attackers at $3000$ m and $6000$ m are able to jam almost all the deployed sensors since they are placed close to each other and they are within the range of the attackers.   For test purposes, we assign almost the  same weight for all three directions of our third objective ( $\approx 0.3$), however, these values can be adjusted by system designers to serve their desires.

\begin{comment}
\begin{table}
  \caption{Fitness Values of Objectives with Random Placement}
  \label{tab:fitness}
  \begin{tabular}{cccccccl}
    \toprule
    Objective Function &1&2&3&4&5&6&7\\
    \midrule
    OF1 & \checkmark &  &  & \checkmark & \checkmark &  & \checkmark\\
    OF2 &  & \checkmark &  & \checkmark &  & \checkmark & \checkmark\\
    OF3 &  &  & \checkmark &  & \checkmark & \checkmark & \checkmark\\
   
  \bottomrule
   Fitness value 1 &  &  &  &  &  &  & \\
   Fitness value 2 &  &  &  &  &  &  & \\
   Fitness value 3 &  &  &  &  &  &  & \\
\end{tabular}
\end{table}

\begin{table}
  \caption{Fitness Values of Objectives with Random Placement of 21 sensors from OpenSky}
  \label{tab:fitness}
  \begin{tabular}{cccl}
    \toprule
    Experiment  & OF1 & OF2 & OF3 \\
    \midrule
    1 & 0.02203632 &  &   \\
    2 &  & 0.02732189 &   \\
    3 &  &  & -0.06136491  \\
    4 & 0.02203632 & 0.02732189 &   \\
    5 & 0.02203632 &  & -0.06136492  \\
    6 &  & 0.02732189 & -0.06136492  \\
    7 & 0.02203632  & 0.02732189 & -0.06136492  \\
  \bottomrule
  
\end{tabular}
\end{table}
\end{comment}

\subsection{\textbf{Results}}

\begin{figure}[tb] 
%  \fbox{
  \begin{subfigure}[b]{0.45\linewidth}
    \centering
     {
        \includegraphics[width=0.95\linewidth]{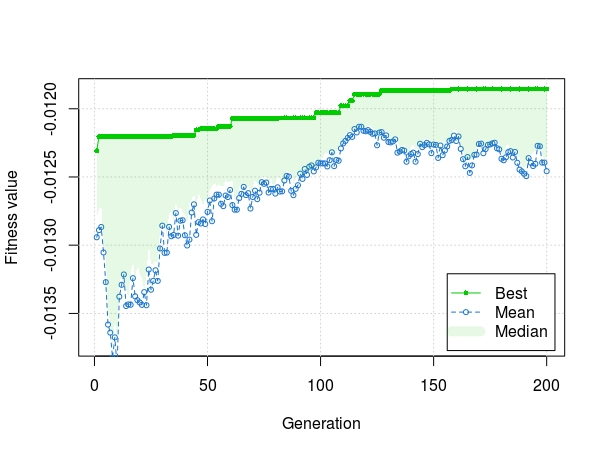} 
    }
    \caption{Objective 1} 
    \label{a} 
  \end{subfigure}%% 
  \begin{subfigure}[b]{0.45\linewidth}
    \centering
     {
        \includegraphics[width=0.95\linewidth]{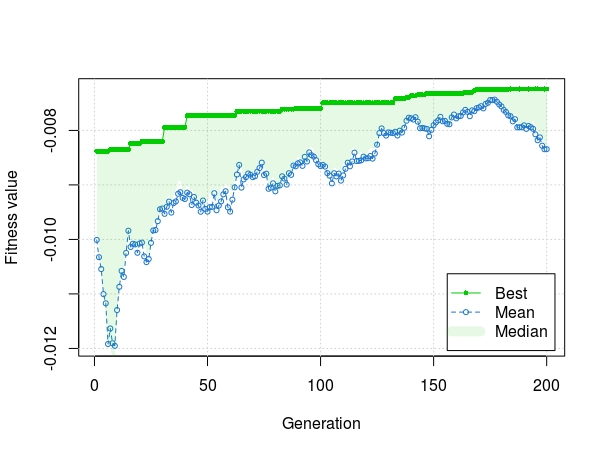} 
    }
    \caption{Objective 2} 
    \label{b} 
  \end{subfigure}%} 
  
%   \fbox{
  \begin{subfigure}[b]{0.45\linewidth}
    \centering
     {
        \includegraphics[width=0.95\linewidth]{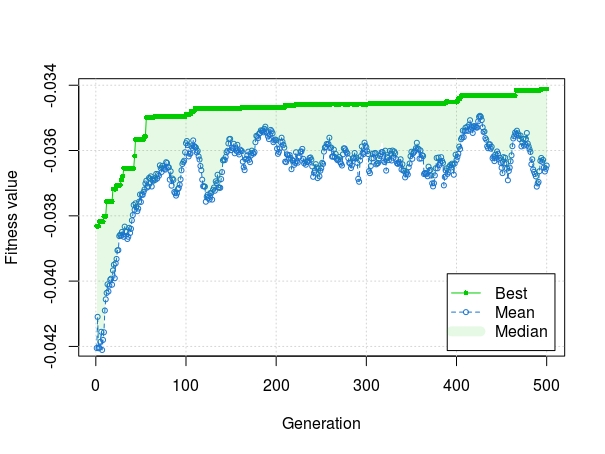} 
    }
    \caption{Objective 3} 
    \label{c} 
  \end{subfigure} %}
  %\vspace*{10 mm}
  \begin{subfigure}[b]{0.45\linewidth}
    \centering
     {
        \includegraphics[width=0.95\linewidth]{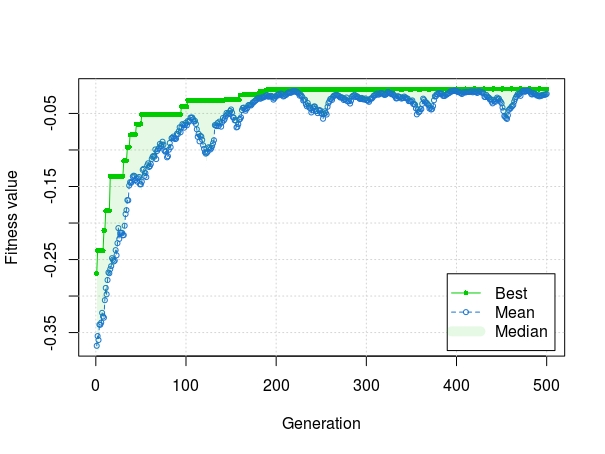} 
    }
    \caption{All objectives together} 
    \label{d} 
  \end{subfigure}
  \caption{The fitness Function of each objective by using the GA algorithm and the fitness function of all objectives together.}
  \label{fig:Fitness_GA} 
  \vspace*{-5mm}

\end{figure}

Now, we are considering the same area of study and with $n$ new sensors that we are wishing to place considering the objective functions that we have.  First of all, we use the basic GA to get the fitness value of each objective alone and then the fitness value of all together across different number of sensors. Figure \ref{fig:Fitness_GA} depicts the fitness function values of the three objectives and the number of generations that are needed to reach the near-optimal solution with $n=30$. The third objective takes more generations to get the near-optimal solution, hence it combines three directions; two maximization problems and one minimization problem. Also, we test how the GA algorithm is able to get a solution of all objects together. Based on the experiments, we observed that at $n=30$ the genetic algorithm stops  improving the fitness value, so it will be best minimum number of sensors that are required to cover the whole area of study in respect to the three objectives that we defined. Thus, we fix the number of sensors to $n=30$ for the rest of experiments.   

\begin{figure}[tbp]
\centering
{\includegraphics[width=.80\linewidth]{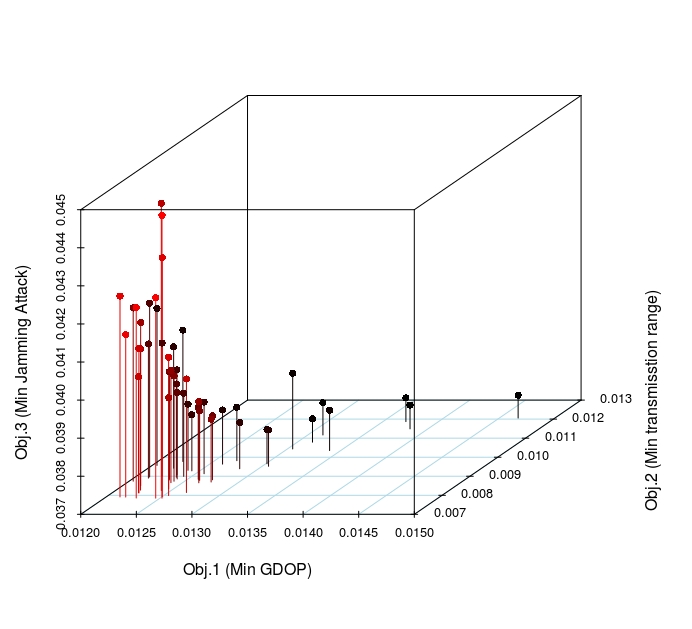}}
\vspace*{-3mm}
\caption{The fitness function values of all objectives for placing $n=30$ new sensors. Each dot represents the solution of a set of locations of $n=30$ ADS-B sensors.}
\label{fig:Fitness_NSGA_all_30}
\vspace*{-3mm}
\end{figure}

The obtained solution by GA considers all objectives together, where the system designers are not able to select the best solution for each objective.  Thus, We run the NSGA-II and we got a set of solutions where each solution is non-dominated by another. Figure \ref{fig:Fitness_NSGA_all_30} shows set of solutions to place  $n=30$ new sensors with their fitness function. We have to illustrate that each solution represents the locations of the $n=30$ sensors that we are wishing to place. The solution with minimal fitness value considers better than the one with high values. As an example, from the figure \ref{fig:Fitness_NSGA_all_30}, we can say the solutions that are located close to the bottom left corner are near-optimal solutions for the first and second objectives, while the solutions are considered near optimal based on the third objective where they are close to the ground surface of the cube.

\begin{figure}[tbp]
\centering
{\includegraphics[width=.80\linewidth]{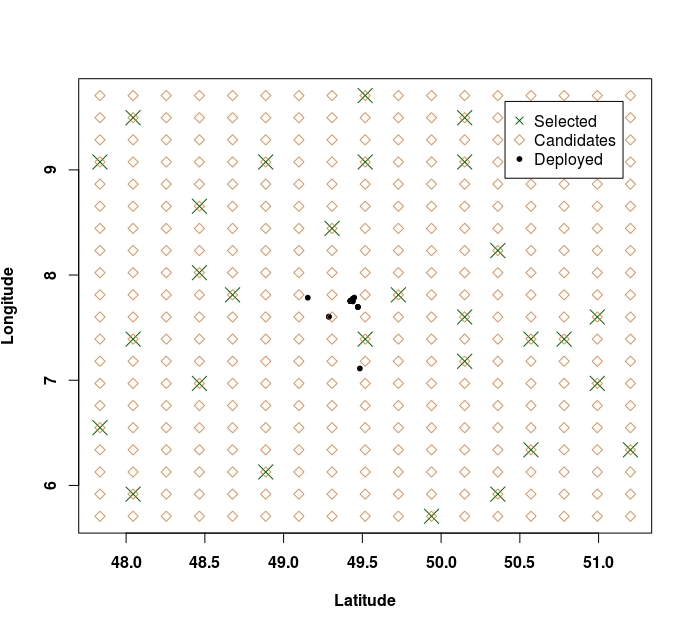}}
\vspace*{-3mm}
\caption{The geographical representation of the $21$ deployed sensors, $30$ new selected sensors, and $400$ candidates locations that the NSGA-II algorithm have to choose from. }
\label{fig:sensors}
\vspace*{-3mm}
\end{figure}

Figure \ref{fig:sensors} shows the location of the already $21$ deployed sensors, $400$ candidates sensors, and the $30$ selected sensors out of $400$. As we can see the deployed ones are concentrated close to each other, while the obtained solution from NSGA-II distributed the sensors over the whole area in away guarantee full coverage and at the same time satisfies the objective functions. 

\begin{figure}[tbp]
\centering
{\includegraphics[width=.80\linewidth]{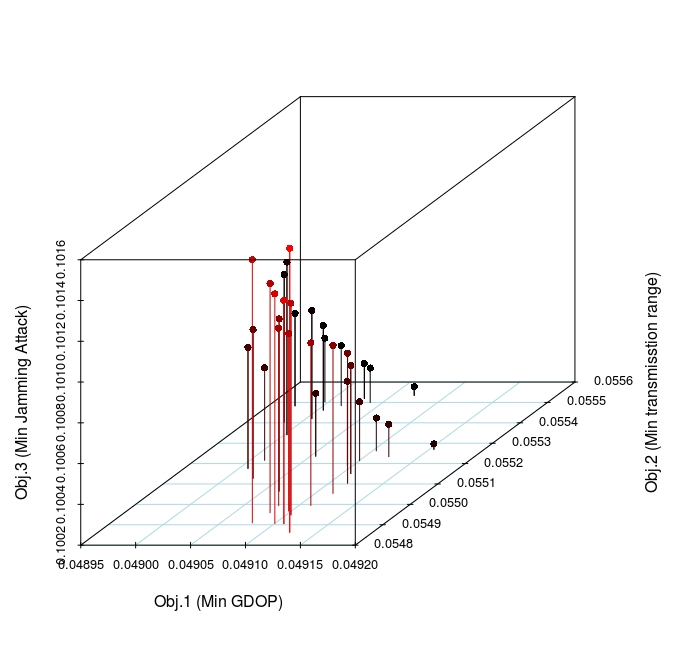}}
\vspace*{-3mm}
\caption{The fitness function values of all objectives for placing $15$ new sensors in addition to $21$ already deployed sensors  $n=15+21$.}
\label{fig:Fitness_NSGA_dep_15}
\vspace*{-3mm}
\end{figure}

Moreover, we consider the scenario where we have to add $n$ new sensors to the already deployed ones to get a near-optimal solution. We assume we have to add $n=15$ new sensors and we need to check the set of solutions to place them with the $21$ sensors from OpenSky. Figure \ref{fig:Fitness_NSGA_dep_15} presents the set of sensors with their fitness values. As shown, the set of solutions go slightly to the right-up which means still we can be close to the optimal scenario with $n=15$ sensors and the solutions could be enhanced if we increase the number of sensors, but we restrict ourselves here to $n=15$ to show how the system designers with low budget can get benefit of our method to place their sensors with the already deployed ones.

\begin{figure}[!t]
\centering
\includegraphics[ width=3.7in]{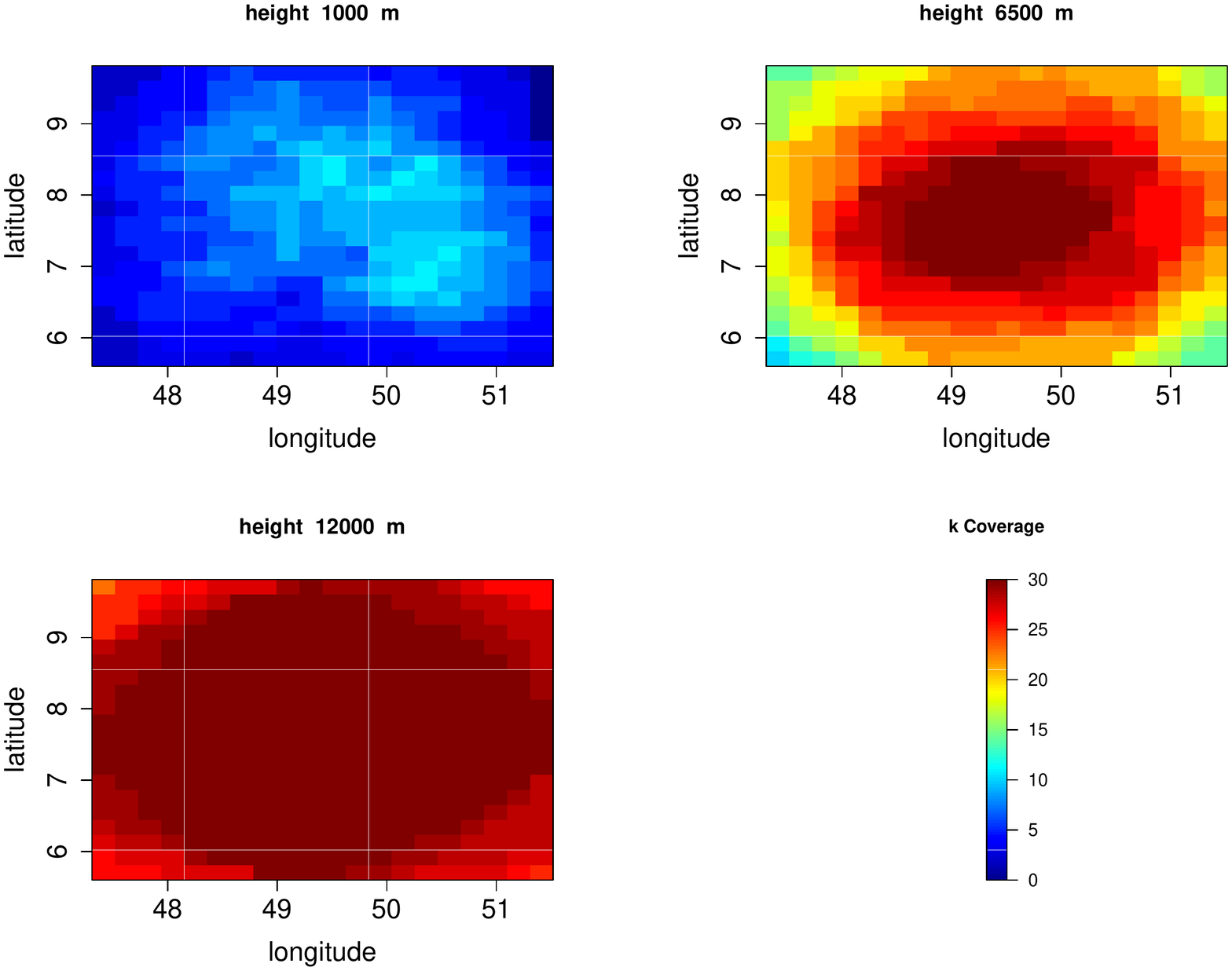}
\caption{Simulated k-coverage heatmap for best placement solution  of $n=30$ ADS-B sensors.}
\label{fig:Kcovbestsol}
\vspace*{-3mm}
\end{figure}

\begin{figure}[!t]
\centering
\includegraphics[ width=3.7in]{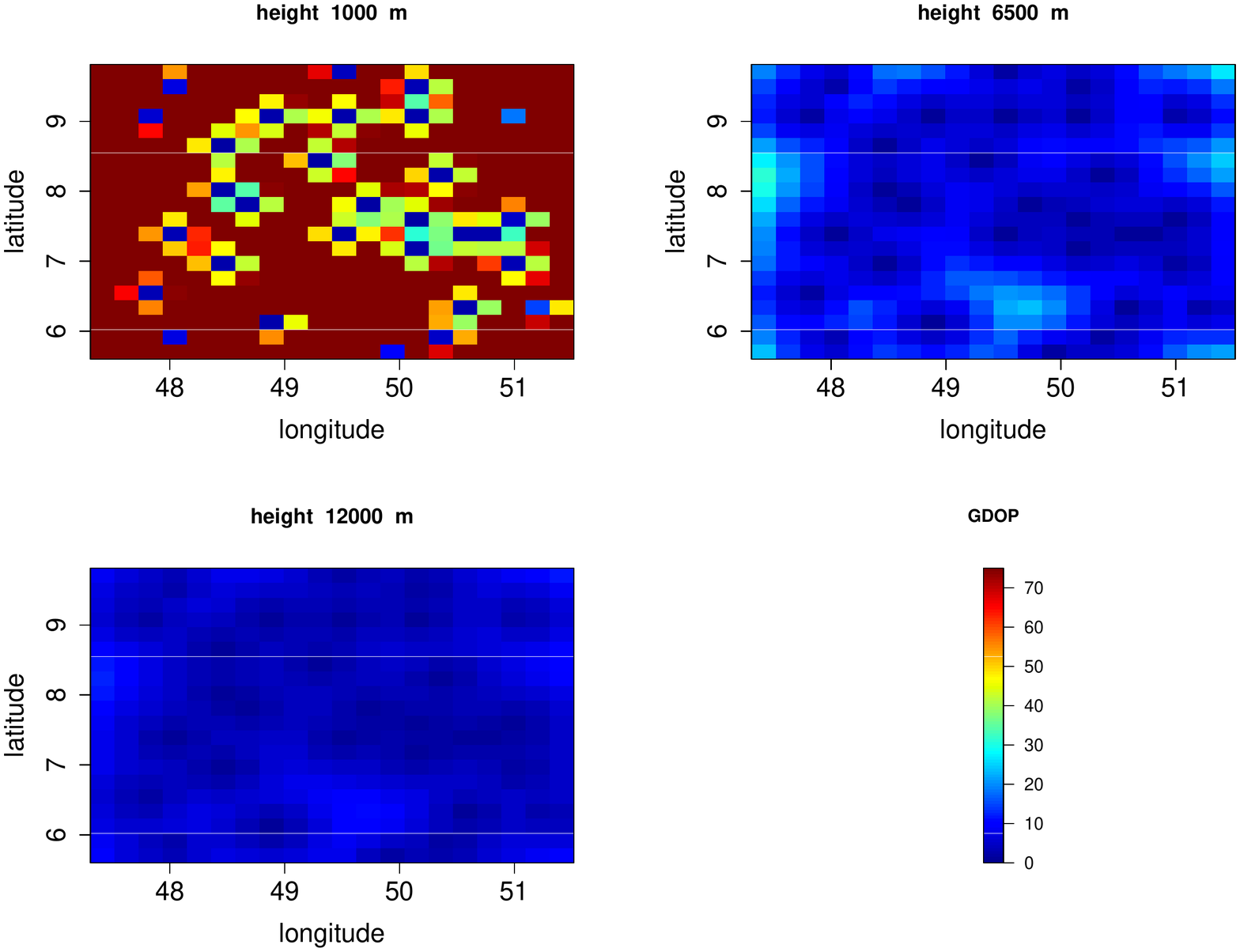}
\caption{Simulated GDOP values for best placement solution of $n=30$ new ADS-B sensors}
\label{fig:GDOPbestsol}
\end{figure}

As we are considering increasing the coverage and reducing the GDOP values, we test the k-coverage and GDOP distribution for one of the obtained solutions from NSGA-II. As we can see from figure \ref{fig:Kcovbestsol} how the selected sensors can still achieve good coverage where MLAT or Lightweight checks can be used to verify the aircraft locations while the GDOP value is reduced significantly as shown in figure \ref{fig:GDOPbestsol}.  %More detail on the percentage of the distribution of GDOP values is presented in Figure \ref{fig:GDOP_Kcove_1} and Figure \ref{fig:GDOP_Kcove_2} in the Appendix \ref{Exp-Appendix}. The figure shows that the GDOP of the current deployment sensors above $60$ is mostly around $90\%$, while it is reduced to only $24 \%$ with the selected ones.  

\begin{figure}[tbp]
\centering
{\includegraphics[width=.80\linewidth]{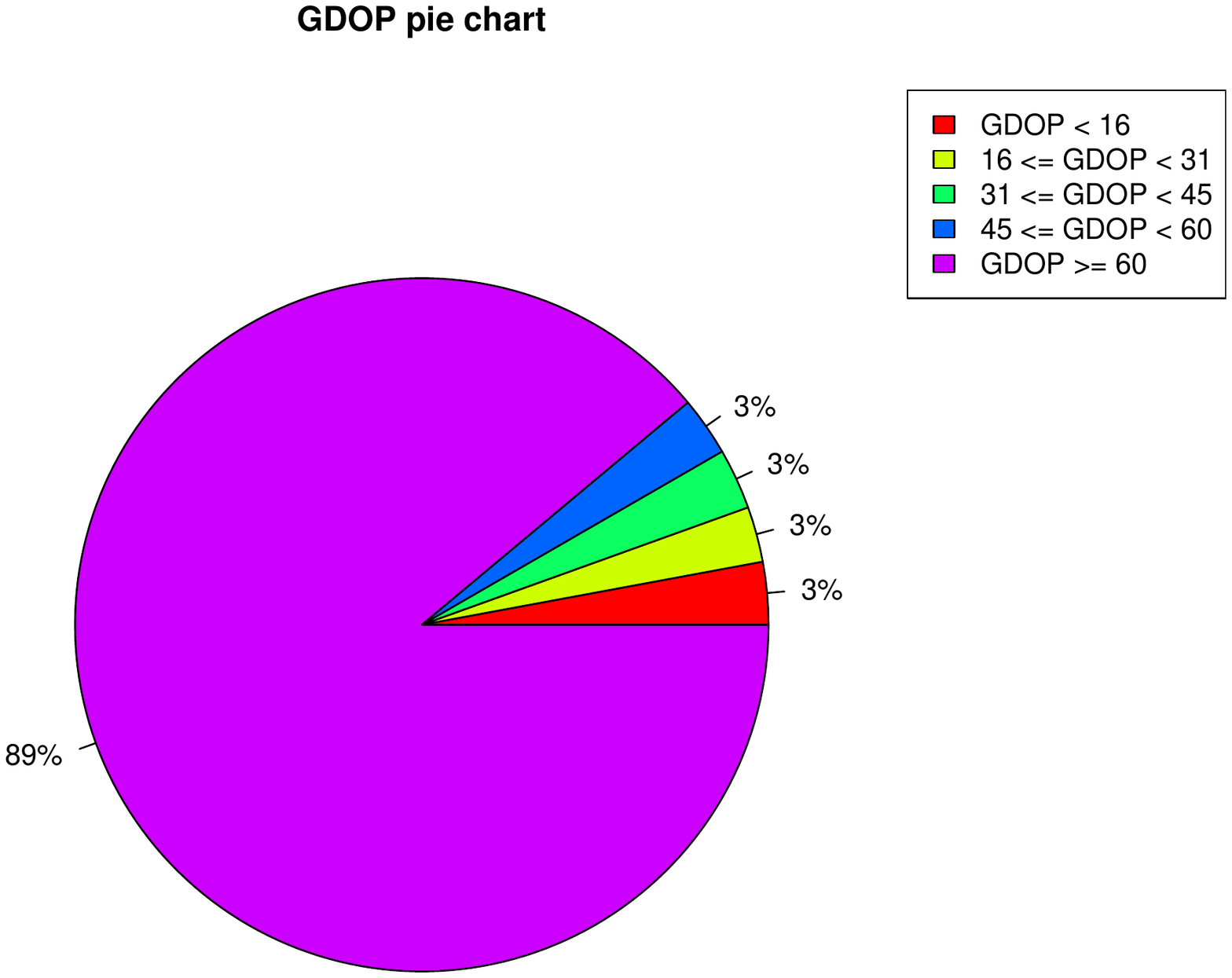}}
\caption{Deployed sensors}
\label{fig:GDOP_Kcove_1}
\vspace*{-3mm}
\end{figure}

\begin{figure}[tbp]
\centering
{\includegraphics[width=.80\linewidth]{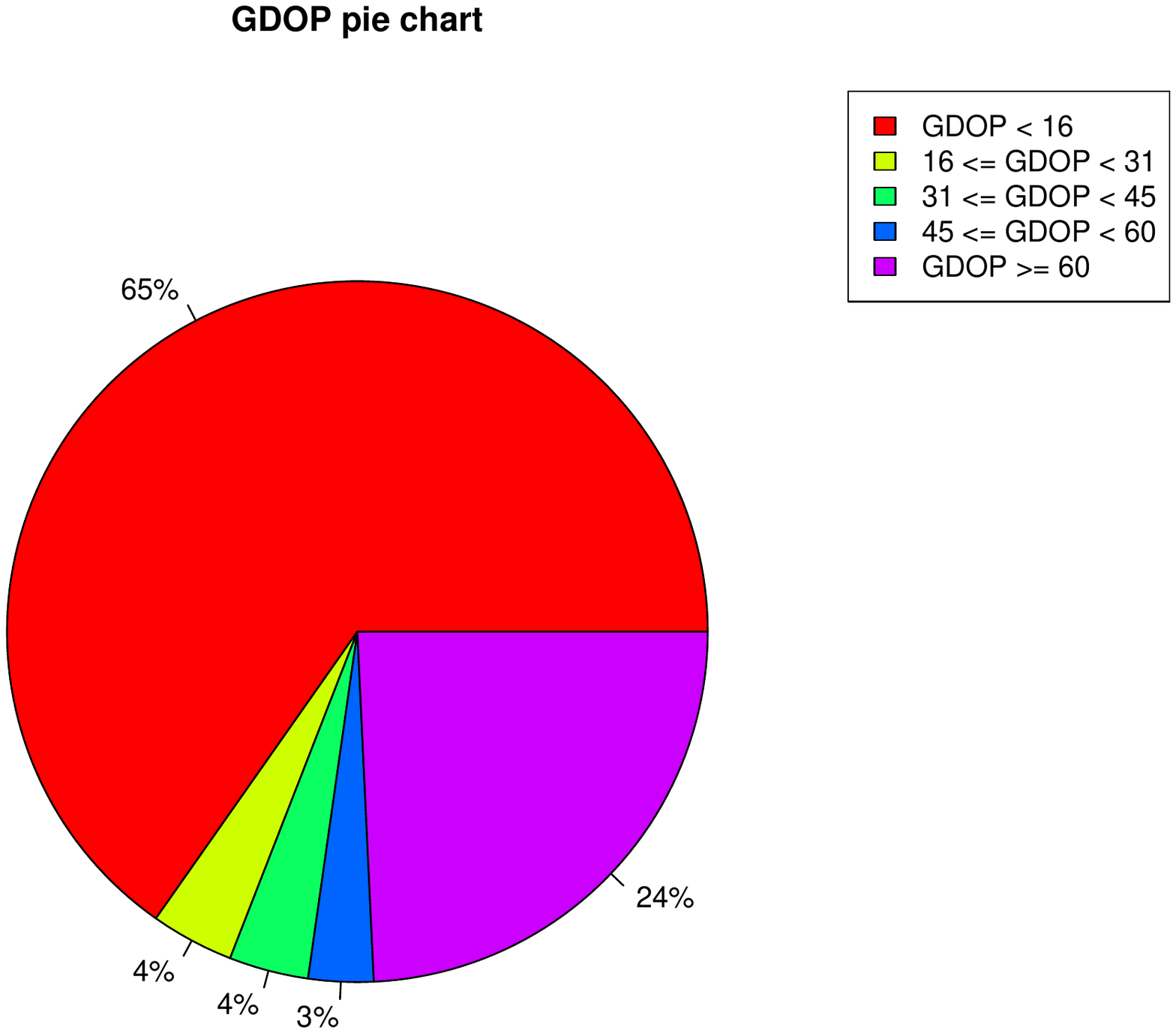}}
\caption{Selected Sensors}
\label{fig:GDOP_Kcove_2}
\vspace*{-3mm}
\end{figure}  
%\end{comment}

The percentage of the distribution of GDOP values is presented in Figure \ref{fig:GDOP_Kcove_1} and Figure \ref{fig:GDOP_Kcove_2} .  The figures shows that the GDOP of the current deployment sensors above $60$ is mostly around $90\%$, while it is reduced to only $24 \%$ with the selected ones. 

\begin{figure}[tbp]
\centering
{\includegraphics[width=.80\linewidth]{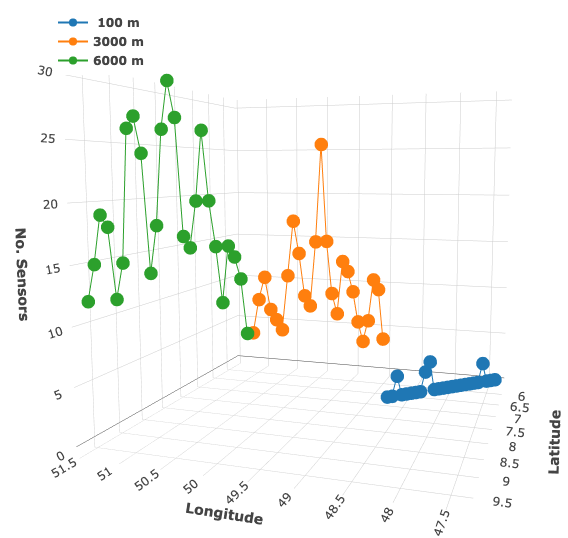}}
\caption{The number of sensors that are affected by the jamming attack setup with  $n=30$ new selected sensors and 75 jammers.}
\label{fig:Jamming_new}
\vspace*{-3mm}
\end{figure}

Lastly, we test how the locations of the selected solution of $n=30$ sensors are resistant to the jamming attack. As we can see from figure \ref{fig:Jamming_new} the number of sensors that are affected by the jamming is reduced compared with the one of the deployed ones. Also, we have to mention that, these numbers are across the whole area, while the ones that are shown in \ref{fig:Jamming_dep} are all concentrated within the range of deployed sensors.  

\section{Related Work}

Aircraft tracking becomes vital with the widespread of cyberattacks. Thus, some existing approaches verify the trustworthiness of these received messages. Multilateration (MLAT)  \cite{Mantilla-Gaviria2015} is one of the most famous approaches that have been used. However, the percentage of messages that can be veritably by MLAT  with GDOP $<10$ is only around 5.24\%  from the whole messages \cite{strohmeier2018k} because such verification requires at least four sensors to receive the message. Another K-NN based approach \cite{strohmeier2018k} is proposed to verify the messages that are received by two sensors. This approach increases the percentage of the messages that can be verified up to 41.48\% but in 2D dimensions. Other solutions \cite{MAVPro} also proposed to verify the messages that are received by one sensor but with less accuracy.  

All of these location checks depend on the number and the location of the receiving sensors. In an unstructured placement of  ADS-B receivers, the location verification checks become inapplicable and the aircraft may not be tracked by the ATC. Recently, the OSP problem in an avionic context has been investigated  \cite{monteiro2015detecting}. The authors disregard the requirement of aircraft height (altitude) verification and they only verify the latitude and longitude, they did that based on the assumption that MLAT  applied with coplanar receivers generally results in a poor vertical dilution of precision.   Such a way could minimize the horizontal error that is computed as the ratio between the common intersection area between all k receiving sensors (also known as the k-coverage or k-intersection area) and the cumulative area that is covered by those receivers. However, this approach is inaccurate for location verification since the aircraft has to be considered.  We believe that, if the receivers are placed and spread carefully, then the coplanarity assumption can be broken due to the Earth curvature.      

In \cite{nijsure2016adaptive} the authors addressed the problem of optimal sensors selections through the aircraft tracking phase. This method assumes the  deployed sensors are thoughtfully placed, thus, if the sensors are poorly chosen then an acceptable GDOP value can be achieved. 

Authors in \cite{Maurosensors} proposed a procedure to place the ADS-B to enhance the mode-S multilateration stations for airport surveillance. Their approach takes into account the LoS, and the GDOP along with Cramer–Rao Lower Bound (CRLB) \cite{ucinski2004optimal} analysis. Their standard was able to enhance the MLAT by using the GA to get the station locations. However, all of the discussed methods were targeting only one objective and non of them targets the jamming attack reduction. Thus, our work addresses the OSP as MOOP and solves it using the NSGA-II that provides non-dominated solutions.

\section{Conclusion}

A Multi-Objective Optimization Problem (MOOP) to place the ADS-B sensors on-ground is proposed by this paper. The Optimal Sensor Placement Problem has been tackled with respect to three objective functions. The first and the second objectives aim to provide an optimal solution that guarantees full coverage where each ASD-B message has to be received by at least one receiver and at the same time allows location checks of the aircraft to verify the trustworthiness of the received claim location in the ADS-B message. While in our third objective, we aim to reduce the effect of jamming attacks by placing sensors in a way where the number of sensors that are affected by the attackers is minimized. We use the Non-dominated Sorting Genetic (NSGA-II) algorithm to optimize and get our set of solutions. The results show, how the obtained solutions are optimized simultaneously and each solution is non-dominated by another which gives the system designer the flexibility to select the best optimal solution based on their budget and needs.

%Aircraft tracking becomes a vital with the widespread of the cyberattacks. The placement of on-ground ADS-B (Automatic Dependent Surveillance–Broadcast) receivers considers crucial  aspect to enable the aircraft to be tracked by the ATC system and at the same time allows the ATC to control the air-traffic and check the trustworthiness of ADS-B messages.   

\section*{Acknowledgments}
 This work is supported by the Center for Cyber Security at New York University Abu Dhabi (NYUAD). The authors gratefully acknowledge financial support of and interaction with armasuisse Science \& Technology. We would like to thank the OpenSky Network for support, more specifically Martin Strohmeier for his collaboration and feedback at the early stages of this work.

%\section*{Acknowledgments}

\bibliographystyle{ACM-Reference-Format}
\bibliography{Sensors_Placement2022}

%\appendix
%\section*{Appendix}

%\begin{comment}

%\begin{comment}

%\end{comment}

%\subsection{\textbf{ Lightweight Location Estimation Principle}}

%\subsection{\textbf{Jamming-to-Signal Ratio (JSR)}}
%\subsection{\textbf{Sensor Deployment Optimality Criteria}}

\end{document}